\documentstyle[11pt,amsfonts,amssymb,amscd]{article}
\begin{document}

\title{Noncommutative Topological Quantum Field Theory, Noncommutative Floer Homology, Noncommutative Hodge Theory}
\author{Ioannis P. \ ZOIS\thanks{izois@ppcr.gr}\\
\\ 
Public Power Corporation, Research and Development Department
\\
3, Kapodistriou Street, GR-153 43 Aghia Paraskevi, Athens, Greece
\\}
\date{}
\maketitle
\begin{abstract}

We present some ideas for a possible Noncommutative Topological Quantum Field Theory (NCTQFT) and Noncommutative 
Floer Homology (NCFH). Our motivation is two-fold and it comes both from physics and mathematics: 
On the one hand we argue that NCTQFT is the correct mathematical framework for a quantum 
field theory of all known interactions in nature (including gravity). On the other hand we hope that a possible NCFH will 
apply to 
practically every 3-manifold (and not only to homology 3-spheres as ordinary Floer Homology currently does). The 
two motivations are closely related since, at least in the commutative case, Floer Homology Groups constitute 
the space of quantum observables of (3+1)-dim Topological Quantum Field Theory.
Towards this goal we present some "Noncommutative" Versions of Hodge Theory for 
noncommutative differentail forms and tangential cohomology for foliations.

Classification: theoretical physics, mathematical physics, geometric topology, differential geometry, quantum 
algebra\\

Keywords: Floer Homology, Noncommutative Geometry, Topological Quantum Field Theory, 3-manifolds, K-Theory, 
Quantum Gravity\\

\end{abstract}

\section{Introduction and Motivation}

This article describes some ideas which emerged during our most recent visit at the IHES a few years ago.
Our motivation is twofold: it comes both from physics and from mathematics.

\subsection{Physical Motivation: Why NCTQFT?}

Why 
should one want a Noncommutative extension of Topological Quantum Field Theory (NCTQFT)?\\
 
We shall argue that NCTQFT should be an adequate framework for a unified quantum field theory incorporating 
all known interactions in nature, including gravity.\\

The cornerstone of quantum theory is the principle of particle-wave duality. Although neither gravitons nor gravitational waves have been 
experimentaly observed until now, most physicists take the point of view that quantum gravity-which is currently an elusive
theory-\emph{should exist;} one of the main arguments in favour of its existence is mathematical consistency 
and it goes back to P.A.M. Dirac: 
let us consider Einstein's classical field equations 
which describe gravity (we assume no cosmological constant and we set the speed of light $c=1$):
$$G_{\mu\nu}=8\pi GT_{\mu\nu}$$
In the above equation, $G$ denotes Newton's constant, $T_{\mu\nu}$ denotes the energy-momentum tensor and 
$G_{\mu\nu}$ denotes the Einstein 
tensor which is equal, by definition, to $G_{\mu\nu}:=R_{\mu\nu}-\frac{1}{2}Rg_{\mu\nu}$, where $g_{\mu\nu}$ is 
the Riemannian metric, $R_{\mu\nu}$ 
is the Ricci curvature tensor and $R$ is the scalar curvature. One can see clearly that the RHS of the above
equation,namely the energy-momentum tensor, contains 
mass and energy coming from the other two interactions in nature; mass for instance (of ordinary matter), 
consists of
fermions (quarks and leptons) and we know that these interactions (strong and electroweak) are quantized and hence
the RHS of the equation contains \emph{quantized quantities}. So for \emph{consistency} of the equations, the 
LHS, which encodes geometry, \textsl{should also be quantized}.\\

[Aside 1: one may argue that the LHS may remain 
classical while the RHS may involve the \textsl{average value} of an operator; however such a theory will not be 
essentially different from classical general relativity and probably not qualified to be called quantum gravity, 
what we 
have in mind is the Ehrenfert Theorem from Quantum Mechanics. We think of the above 
field equations as describing, in the quantum level, an actual equality between operators].\\

There are a number of other reasons why physicists would like quantum gravity to exists like the 
elimination of spacetime singularities etc.\\
 
Now the famous and very well-known \emph{Holography Principle}, which has attracted
a lot of attention since 1993 when it was proposed originally by G. 't Hooft (see \cite{thooft}), states that 
quantum gravity should
be a \emph{topological quantum field theory} as defined by Atiyah in \cite{atiyah}. There has been strong evidence 
from the GEO 600 experiment towards the validity of holography (see \cite{geo600}). In fact 
quantum gravity should be a topological quantum field theory even without holography: 
given (for simplicity) a closed Riemannian 4-manifold and the Einstein-Hilbert action which 
contains 
the square root of the scalar curvature of the metric as Lagrangian density, in order to compute the partition 
function 
of the theory
one would have to integrate over all metrics. It is clear that if one was able to perform this functional 
integral, 
the result should be a topological invariant of the underlying manifold simply because "there is nothing else 
left" apart 
from the topology of the Riemannian manifold. We take for brevity the 4-manifold
 to be closed, so Atiyah's axioms for a topological quantum field theory will reduce to obtaining numerical 
invariants 
and not elements of a vector space associated to the boundary (eg Floer Homology Groups of the boundary 3-manifold).
But here there is an impotant question: the partition
function of the Einstein-Hilbert action on a Riemannian manifold should be a topological invariant, but should it
be a \emph{diffeomorphism} \textsl{or a} \emph{homeomorhism} (or even homotopy) invariant? We know from the 
stunning work 
of S.K. Donaldson in the '80s (see \cite{donald}) that the DIFF and the TOP categories \textsl{in dimension 4} 
are two entirely different 
worlds (existence of "exotic" ${\bf R}^{4}$'s). So particularly for the case of 4-manifolds (which is our 
intuitive idea for 
spacetime, at least macroscopically) this question is crucial. 

[Aside 2: We would like to make a remark here: in physics 
literature the term "topological"
really  means "metric independent" without further specification but for 4-dim geometry, this point is 
particularly important].

We do not have a definite answer on this but it is an issue which in most cases it is not addressed to in the 
physics literature;
 however we feel that for quantum physics TOP should be more appropriate as the working category since for example 
in quantum 
mechanics (solutions of Schrodinger equation) one requires only continuity and not smoothness of solutions at 
points 
connecting different regions.\\

But this is not enough; if we want a unifying theory of all interactions, we must have other fields present apart 
from the 
metric (eg gauge fields for electroweak and strong interactions and/or matter fields). We know from the case of the 
\textsl{Quantum Hall Effect, QHE for short} and Bellissard's
work (and others') (see eg \cite{connes}) that the existence of \emph{external fields} "make things 
\emph{noncommutative}". 
For the
particular case of the QHE the presence of a uniform magnetic field turns the
 Brilluin zone of a periodic crystal from a 2-torus to a noncommutative 2-torus (see \cite{connes}). 
Moreover as Connes et al. have shown recently (see \cite{connes2}), 
the full 
4-dim standard model Lagrangian of electroweak and strong interactions with Yukawa couplings and neutrino mixing can be 
geometrically interpreted as the fundamental K-Homology class of a noncommutative manifold arising as the discrete 
product of a spin 4-dim Riemannian manifold with a discrete space of metric dimension $0$ and KO-dimension 6 mod 8. 
Further evidence for this phenomenon,
namely the appearence of noncommutative spaces when external fields are present, comes from string theory: the 
Connes-Douglas-Schwarcz article (\cite{cds}) indicates that when a constant 3-form $C$ (acting as a potential) 
of D=11 supergravity is
turned on, M-theory admits additional compactifications on noncommutative tori. Also in string theory, the 
Seiberg-Witten
article (see \cite{sw}) also discusses noncommutative effects on open strings arising from a nonzero $B$-field. 
So we 
believe there is good motivation
to try to see what a possible \emph{noncommutative topological quantum field theory} should look like since from 
what
we mentioned above, it is reasonable to expect that a unifying quantum theory should have some noncommutativity 
arising from
the extra gauge or other fields present; it should also be a topological quantum field theory since it should 
contain 
quantum gravity.\\

\subsection{Mathematical motivation: Why NCFH?}

We would like to deepen our understanding on 3-manifolds. Floer Homology is a very useful device since it is the 
only known 
homology theory which is only \emph{homeomorphism} and \textsl{not homotopy} invariant. (This distinction lies at the heart 
of manifold topology and it captures the essence of the Poincare conjecture). Yet computations are 
particularly 
hard and the theory itself is very complicated; moreover the notorius reducible connections make things even worse 
and at the end Floer Homology Groups are defined only for homology 3-spheres. We would like to have a hopefully 
simpler 
theory which would apply to a larger class of 3-manifolds. We shall elaborate more on this in the next sections.\\
Let us start by recalling some well-known facts from 3-manifold topology: we fix a nice Lie group $G$, 
say $G=SU(2)$; if $M$ is a 3-manifold with fundamental group $\pi _{1}(M)$, 
then the set 
$$R(M):=Hom(\pi _{1}(M),G)/ad(G)$$ 
consisting of equivalence classes
 of representations of the fundamental group $\pi _{1}(M)$ of $M$ onto the Lie 
group $G$ modulo conjugation \textsl{tends to be discrete}.
 If $M$ is a homology  3-sphere, ie $H_{1}(M;{\bf Z})=0$, (this is a safficient
 condition but not in any way necessary), then $R(M)$ has a 
\emph{finite} number of elements and the \textsl{trivial representation} is 
\textsl{isolated}.\\

There is a well-known 1:1 correspondence between the elements of the set $R(M)$
 and elements of the set\\ 

$A(M):=\{$\emph{flat} $G$-connections on $M\}$/(gauge equivalence)\\

The bijection is nothing other than the \emph{holonomy} of the flat 
connections.\\

 Although $R(M)$ \emph{depends on the homotopy type of} $M$, we can get 
\emph{topological invariants} of $M$, ie invariants under 
\textsl{homoeomorphisms}, if we use the moduli space $A(M)$: depending on how we
 ``decorate'' the elements of $A(M)$, namely by giving different ``labels'' to 
the elements of $A(M)$, we can get the following \emph{topological invariants} 
for the 3-manifold $M$:\\

{\bf 1.} \emph{The (semi-classical limit of the) Jones-Witten invariant}.\\
Pick $G=O(n)$ and for each (gauge equivalence class of) flat 
$O(n)$-connection $a$ say on $M$, we have a flat $O(n)$-bundle $E$ over $M$ 
with flat $O(n)$-connection $a$ along with its exterior covariant derivative 
denoted $d_{a}$; now since $a$ is flat, $d_{a}^{2}=0$ and hence we can form the
  \textsl{twisted de Rham complex} of $M$ by the flat connection $a$ denoted
$(\Omega ^{*}(M,E), d_{a})$, where $\Omega ^{*}(M,E)$ denotes smooth 
$E$-valued 
differential forms on $M$. If we equip $M$ with a Riemanian metric then we can define a Hodge star operator $*$ 
and thus we can also 
define the adjoint operator $d^{*}_{a}$ of $d_a$ which is equal to 
$$d_{a}^{*}=(-1)^{kn+n+1}*d_{a}*$$
(acting on $k$-forms on an $n$-dim Riemannian manifold)
and then finally one can define the 
\textsl{twisted Laplace operator} by the flat connection $a$ to be:
$\Delta _{a}:=d_{a}^{*}d_{a}+d_{a}d_{a}^{*}$.
Then the
\emph{Ray-Singer analytic 
torsion} $T(M,a)$ is a \emph{non-negative real number} defined by the formula (see \cite{rs}):

$$log[T(M,a)]:=\frac{1}{2}\sum_{i=0}^{3}(-1)^{i}i\zeta '_{\Delta _{i,a}}(0)$$

where $\Delta _{i,a}$ denotes the twisted Laplace operator acting on $i$-forms
and

$$\zeta '_{\Delta _{i,a}}(0):=-\frac{d}{ds}\zeta _{\Delta _{i,a}}|_{s=0}=
log D(\Delta _{i,a})$$

and where we call $D(\Delta _{i,a})$ the $\zeta$-function regularised 
determinant of the Laplace operator $\Delta _{i,a}$ (this is a generalisation 
of the logarithm of the determinant of a self-adjoint operator).\\
The 
$\zeta $-function of the Laplace operator $\zeta _{\Delta _{i}}$
is by definition (for $s\in {\bf C}$):

$$\zeta _{\Delta _{i}}(s):=\sum _{\{\lambda _{n}\geq 0\}}\lambda _{n}^{-s}=
\frac{1}{\Gamma (s)}\int_{0}^{\infty}t^{s-1}Tr(e^{-t\Delta _{i}})dt$$

for $Re(s)$ large. Then $\zeta _{\Delta _{i}}$ extends to a meromorphic 
function of $s$ which is analytic at $s=0$.\\

One can prove that \textsl{the Ray-Singer analytic torsion is independent of the Riemannian metric
 if the twisted de Rham cohomology groups are trivial.}\\

If $M$ is a homology 3-sphere (or any other 3-manifold such that the set $A(M)$
 has finite cardinality), then if we sum-up the Ray-Singer analytic 
torsions of all the flat connections 
(since these are finite in number we know the sum will converge), what we shall get 
as a result is a topological invariant of the 3-manifold which is closely 
related to the ``low energy limit'' (or the semi-classical limit) of the 
\emph{Jones-Witten} (or Reshetikin-Turaev) quantum invariants for 3-manifolds (see \cite{witten}). 
More precisely
 the low energy limit of the Jones-Witten quantum invariants for homology 
3-spheres is a finite 
sum of combinations of the Ray-Singer torsions with the corresponding 
Chern-Simons numbers (ie the integral of the Chern-Simons 3-form over the 
compact 3-manifold $M$) of the flat connections.\\

{\bf 2.} \emph{The Casson invariant}.\\
Let $M$ be a homology 3-sphere and pick $G=SU(2)$. If we choose a 
Hegaard splitting on $M$, then assuming that $R(M)$ is regular (ie that the 1st
 twisted de Rham cohomology groups vanish for all flat connections), then each 
element of $R(M)$ aquires an orientation,
namely a ``label'' +1 or -1. Let us denote by $c_{-}$ (resp $c_{+}$) the number
 of elements of $R(M)$ with orientation -1 (resp +1). Both $c_{-}$ and $c_{+}$ 
depend on the Hegaard splitting chosen but their 
\emph{difference} $c:=c_{-}-c_{+}$ \emph{does not} (in fact it behaves like an index) and this integer $c$ is the 
\emph{Casson invariant} of the 3-manifold $M$. Clearly $c$ is well defined 
since the cardinality of $R(M)$ is \emph{finite} and hence both $c_{-}$ and
$c_{+}$ are finite.\\

{\bf 3.} \emph{Floer Homology Groups}.\\
Again $M$ is a homology 3-sphere (and hence both $R(M)$ and $A(M)$ 
have a finite number of elements); we pick $G=SU(2)$,  we denote by $B(M)$ 
the space of \textsl{all} $SU(2)$-connections on $M$ modulo gauge 
transformations and we denote by $B^{*}(M)$ the \textsl{irreducible} ones (a 
connection is irreducible if its stabiliser equals the centre of $SU(2)$ where 
the stabiliser is the centraliser of the holonomy group of a connection). We 
want to do \textsl{Morse Theory} on the $\infty$-dim Banach manifold $B(M)$:\\ 
(i). We find a suitable ``Morse function'' $I:B^{*}(M)\rightarrow {\bf R}$: 
this is
 the integral over $M$ of the Chern-Simons 3-form
$$I(A)=\frac{1}{8\pi ^{2}}\int_{M}Tr(A\wedge dA+\frac{2}{3}A\wedge A\wedge A)$$
 with a \emph{finite} number of \textsl{critical points}; these are precisely 
the  elements of $A(M)$. This is true since the solutions of the Euler-Lagrange
 equations for the Chern-Simons action are the flat connection 1-forms.\\
(ii). Then each element of $A(M)$ aquires a 
``label'' which is the \emph{Morse index} of the critical point; in ordinary 
finite dim Morse theory  this is equal to the number of negative eigenvalues of
 the Hessian. But the Hessian of the Chern-Simons function is unbounded below 
and we get $\infty$ as Morse index for every critical point. So naive 
immitation of ordinary finite dim Morse theory techniques do not work.\\

Floer in \cite{floer} observed the following crucial fact: if we pick a Riemannian metric on $M$, then considering
 the noncompact 4-manifold ${\bf R}\times M$ along with its 
corresponding Riemannian metric, a continuous 1-parameter family of connections
 $A_{t}$ on $M$ corresponds to a unique connection ${\bf A}$ on 
${\bf R}\times M$; then, choosing the axial gauge (0th component of the 
connection vanishes), the \textsl{gradient flow} equation 
for the Chern-Simons function $I$ on $M$ corresponds to the \textsl{instanton 
equation} on the noncompact 4-manifold ${\bf R}\times M$:
$$\partial _{t}A_{t}=*F_{A_{t}}\Leftrightarrow F_{{\bf A}}^{+}=0.$$
 Then consider the \textsl{linearised} instanton equation $d_{{\bf A}}a=0$, 
where $a$ is a small perturbation. This operator is not elliptic; we perturb it
to $D_{{\bf A}}=-d_{{\bf A}}^{*}\oplus d_{{\bf A}}^{+}$ to make it elliptic. 
Then the \emph{finite} integer Morse index for each critical point comes as 
the relative (with respect to the trivial flat connection) Fredholm index of 
the perturbed elliptic operator $D_{{\bf A}}$. In this way the moduli space 
$A(M)$ aquires a ${\bf Z}/8$
grading and then we follow ideas from Morse theory:\\ 
(iii). We define the Floer-Morse 
complex using as generators the critical points and the ``differential'' is 
essentially defined by the flow lines of the critical points. Taking the 
cohomology in the usual way we get the \emph{Floer homology groups} of $M$. 
The Euler characteristic of the 
Floer-Morse complex equals twice the Casson invariant (see \cite{donald}).\\

{\bf Remarks:}

(a). The structure Lie group $SU(2)$ can be replaced by another group, say 
$U(2)$.\\
(b). We assumed that all critical points were not only \emph{non-degenerate} 
(i.e. $H^{1}_{A}(M)=0$, this denotes the first twisted de Rham cohomology group
 of $M$ by the flat connection $A$), but in fact \emph{acyclic} (i.e. 
$H^{0}_{A}(M)=H^{1}_{A}(M)=0$). If this is not the case, then the theory just 
becomes more complicated and one has to use \emph{weighted spaces}.\\
(c). One needs a restriction of the form $b^{+}>1$ in order to be able to prove
 independence on the choice of the Riemannian metric (the Riemannian metric 
defines a Hodge star operator whose square equals $1$, hence its eigenvalues 
are $\pm1$; this gives a splittting of the space of 2-forms into positive and 
negative eigenspaces and $b^{+}$ simply denotes the \textsl{positive} part of 
the 2nd Betti number).\\
(d). Reducible connections create more severe problems; this is the main reason
 why people usually work with homology 3-spheres:  apart from having a finite 
number of gauge equivalence classes of flat connections, they have a unique 
reducible connection which is the trivial flat connection which is moreover 
isolated. If one wants to take the reducible connections into account as well, 
then one has to use \emph{equivariant} Floer homology. This is a lot more 
complicated and less satisfactory as a theory since equivariant Floer Homology 
groups may be \textsl{infinite dimensional} and hence there is no Euler 
characteristic for the equivariant Morse-Floer complex; also there is no 
Casson invariant known in this case.\\

 All the above depend crucially on the fact that $R(M)$ (or equivalently 
$A(M)$) has finite cardinality; the most convenient case that this is 
guaranteed is if $M$ is a 
homology 3-sphere. So the question is: what happens if $M$ is such that $R(M)$ 
does not have finite cardinality? Is there a chance to define the 
analogue of the Casson invariant say in this case or even more than that, a 
Floer homology?\\

We believe \textsl{``yes''} and this is precisely the point we are trying to 
develope here.\\

The key idea is the following: we want to replace $R(M)$
by another more stable and better behaving moduli spcace. To do that we use 
as our basis a recent result by David Gabai (see \cite{gabai}): For practically
 any 3-manifold 
$M$ (closed, oriented and connected), the moduli space $N(M)$ of taut codim-1 
foliations modulo coarse isotopy has \emph{finite} cardinality.\\

More concretely: a codim-1 foliation $F$ on a given manifold $M$ is given by an integrable subbundle 
$F$ of the tangent 
bundle $TM$ of our 3-manifold $M$. A codim-1 foliation $F$ on $M$ is called \textsl{topologically taut} if 
there exists
a circle $S^{1}$ which intersects transversely all leaves. A codim-1 foliation 
is called \textsl{geometrically taut} if there exists a Riemannian metric on 
$M$ for which all leaves are minimal surfaces (ie they have mean curvature 
zero). One can prove that a codim-1 foliation is geometrically taut if and only
 if it is topologically taut. Foliations in general are very flexible 
structures and the taut foliations are the most rigid ones. Let us call the 
quotient bundle $Q:=TM/F$ the \textsl{transverse bundle} to our foliation.\\

Let $M$ be a Riemannian 3-manifold. Two codim-1 foliations on $M$ are called
\emph{coarse isotopic} if up to isotopy of each one of them their oriented 
tangent planes differ pointwise by angles less than $\pi $. Then Gabai proves 
the following (Theorem 6.15 in \cite{gabai}): Given any closed, orientable, 
atoroidal
3-manifold $M$ with a triangulation, there exists a finite non-negative integer
$n(M)$ such that any taut codim-1 foliation on $M$ is coarse isotopic to one
of the $n(M)$ taut codim-1 foliations. The condition that $M$ should be 
atoroidal may be relaxed as Gabai points out. It is clear that $n(M)$ is the 
cardinality of the Gabai moduli space $N(M)$.\\

The crucial fact is that although the definition of coarse isotopy depends on 
the Riemannian metric, the number $n(M)$ \emph{does not.}\\

Let us emphasise here that although the Gabai moduli space is finite 
practically for any 3-manifold, it may turn out to be \emph{empty} [for example, $S^{3}$ has no taut codim-1 
foliations].\\

The key idea then is to try to mimic the constructions of the (commutative) topological invariants described 
above (Ray-Singer torsion, Casson, Floer homology groups) by replacing the moduli space of flat connections 
modulo gauge with taut codim-1 foliations modulo coarse isotopy. From the moment that foliations enter the scene, 
noncommutative geometry becomes relevant since it can supply a wealth of new mathematical tools. This means 
that in principle one could use noncommutative geometric tools to define new invariants for ordinary (commutative)
manifolds. But this is not the end of the story: One might even also try to use noncommutative tools and the 
aforementioned strategies in order to define topological invariants for noncommutative spaces (noncommutative 
manifolds).\\

\section{Available mathematical tools to study and classify foliations}

As it is clear form our previous discussion, (commutative) topological invariants for 3-manifolds are constructed by giving various 
"labels" to gauge classes of connection 1-forms. Following the same strategy then, the next order of business is to find ways to 
"decorate" or "label" (coarse isotopy classses of) taut codim-1 foliations. What are the known topological invariants for foliations?\\

[Aside 3: It seems that the simplest commutative invariant is the Casson invariant, so the simplest idea would be 
to try to see if one can immitate the definition of the 
Casson invariant using the Gabai moduli space. Namely if one chooses a Hegaard 
splitting, can one define a Casson type of invariant by giving "orientations" to taut codim-1 foliations?
We have no definite answer to this question].\\

\subsection{Foliation invariants}

Coming back to the topological invariants for foliations, the first we encounter is the Godbillon-Vey invariant 
(see for example \cite{candel}): This is the integral over our 
compact 3-manifold $M$ of the
Godbillon-Vey class which for codim-1 foliations on $M$ is a 3-dim real de Rham
 cohomology class defined as follows: Suppose that the foliation on $M$ is defined via a transversely 
oriented, 
codim-1, integrable subbundle $F$ of the tangent bundle $TM$ of our closed, oriented and
connected 3-manifold $M$. Locally $F$ is defined by a nonsingular 1-form say 
$\omega$ where $F$ consists precisely of the vector fields which vanish on 
$\omega$ (ie the fibre $F_{x}$ where $x\in M$ equals Ker $\omega _{x}$). The 
integrability condition of $F$ means that $\omega\wedge d\omega =0$. This is 
equivalent to $d\omega=\theta\wedge \omega$ for another 1-form $\theta$. Then 
the Godbillon-Vey class is the 3-dim real de Rham cohomology class 
$[\theta\wedge d\theta ] \in H^{3}(M;{\bf R})$. The problem however with the GV 
invariant is that it is only invariant 
under \textsl{foliation cobordisms} (see \cite{candel}) which is a 
\emph{more narrow} equivalence relation than 
coarse isotopy, hence we may lose the finiteness of the Gabai moduli space 
(equivalently if we use the GV-invariant, we should restrict ourselves to only 
those 3-manifolds with a finite number of taut codim-1 foliations modulo 
foliation cobordisms).\\ 
  
A possibly useful second foliation invariant is the invariant for foliated manifolds 
that the author intruduced some years ago (see \cite{zois}) using indeed 
noncommutative geometry tools, in particular Connes' pairing between cyclic 
cohomology and K-Theory. The foliation has to be transversely oriented with a holonomy invariant transverse measure, 
these restrictions are quite mild. 
Connes' approach to foliations as described in 
\cite{connes} is to complete the 
holonomy groupoid of a foliation to a $C^{*}$-algebra and then study its 
corresponding K-Theory and cyclic cohomology. The invariant in \cite{zois} is 
constructed by defining a canonical K-class in the K-Theory of the foliation 
$C^{*}$-algebra and then pair it with the \textsl{transverse fundamental cyclic
 cocycle} of the foliation. To give a flavour of what that means we describe it
 in the commutative case, ie when the foliation is a fibration, in particular a principal $G$-bundle 
(where $G$ is a nice Lie group): if we have a 
fibration seen as a foliation over a compact manifold (the foliated manifold is
 the \textsl{total space} of the fibre bundle), 
then this transverse fundamental cyclic cocycle is the fundamental homology 
class of the base manifold which is transverse to the leaves=fibres; the 
$C^{*}$-algebra is Morita equivalent to the commutative algebra of functions on
 the base manifold. By the Serre-Swan theorem the K-Theory of this commutative 
algebra coincides with the Atiyah topological K-Theory of the base manifold and
 Connes' pairing reduces to evaluating say Chern classes over the fundamental 
homology class of the base manifold (here we use the 
Chern-Weil theory to go from K-Theory to the de Rham cohomology). The key 
property of the canonical K-class constructed in \cite{zois} is that it takes 
into account the natural action of the holonomy groupoid onto the transverse 
bundle of the foliation. 

[Aside 4: In some sense this class is similar to the canonical class in $G$-equivariant 
K-Theory,
 for $G$ 
some Lie group acting freely on a manifold, the situation is more complicated in the foliation case since instead of a 
Lie group we have the holonomy groupoid of the foliation acting naturally on the transverse bundle].

We also need the result that the $G$-equivariant 
K-Theory of the total space of the principal $G$-bundle is isomorphic to the 
topological K-Theory of the quotient by the group action (since this is a 
$G$-bundle, the quotient by the $G$-action is the base manifold). But this 
invariant has not yet 
been properly understood: obviously if it is to be used to define invariants 
for 3-manifolds using the Gabai moduli space it should be invariant under 
coarse isotopy or under a broader equivalence relation.  For the 
moment this point is unclear.\\

The Heitsch-Lazarov analytic torsion in \cite{heitsch} is defined for foliated flat bundles and it does not seem to 
be of any use 
here since it is exactly the flat connections moduli space which we want to replace.\\

A third possibility for a new topological invariant for foliations  which seems interesting, following what we know from the 
commutative case, is to try to define a \emph{Ray-Singer torsion for foliated 
manifolds} and then try to see if this is invariant under coarse isotopy. In order to define the Ray-Singer 
analytic tosrion one needs a flat connection. For the case of foliations, a flat connection always exists, it is 
our friend the 1-form $\theta$ appearing in the definition of the GV-class; this can indeed be seen in a natural way as
a connection on the transverse bundle (for arbitrary codimension $q$, $\theta$  can be seen as a flat connection on the 
$q$th exterior power of the transverse bundle, this is always a line bundle). This 1-form is sometimes refered to as the
(partial) \emph{flat} Bott connection; it is flat (=closed since this is real valued 
ie Abelian),
 only when restricted to the leaf directions (which justifies the term partial; this is harmless, it can be extended
to a full connection by, for example, using a Riemannian metric).\\

\subsection{(Co)homology theories for foliations}

The next level of complication is the use of more refined tools to study foliations than numerical invariants 
and this is (co)homology theory. In general, there are three known cohomology theories which can be used to study 
foliations: the tangential cohomology, the Hochschild (and cyclic) cohomology of the corresponding foliation 
$C^{*}$-algebra and the so-called Haefliger cohomology (which has been used in
 the construction of the Heitsch-Lazarov analytic torsion).\\
 
For the definition of the corresponding foliation algebra and Hochschild and cyclic cohomology of the corresponding foliation algebra, 
one can see 
\cite{connes}.\\

Perhaps the easiest way to describe tangential cohomology is to start thinking of foliations as generalisations 
of flat vector bundles. Following these lines, one way to manifest the integrability of a flat connection $a$ 
say is
to point out that its exterior covariant derivative $d_{a}$ has square zero 
$d_{a}^{2}=0$, ie it is a differential. Something similar happens for 
foliations if one considers the ``tangential'' (or ``leafwise'') exterior 
derivative on the foliated manifold which is taking derivatives along the leaf 
directions only; the integrability condition means that the leafwise (or tangentail) exterior derivative has 
square zero, hence it is a differentail and this in turn enables one to define the 
``tangential Laplace operator'' 
along with the so called \emph{tangential cohomology} 
and it has corresponding \emph{tangential Chern classes} (see \cite{moore}) by using a Riemannian metric and following the same 
strategy as one does for ordinary de Rham cohomology. 
In the above sense tangential cohomology can be seen somehow as a 
generalisation of the twisted de Rham cohomology by a flat connection. Under 
the light of this note the analytic torsion defined by Heitsch-Lazarov in 
\cite{heitsch} has some unsatisfactory properties for our purpose since it is 
a torsion for a foliated flat bundle (namely a flat bundle whose base sapce is
 in addition, foliated, and so the total space carries essentially 3 structures
: the fibration, the foliation where the leaves are covering spaces of the base
 space--flatness--and another foliation which under the bundle projection 
projects leaves to leaves. One of the main points in this article is to develop a Hodge theory for tangential 
cohomology. This will be done in section 4 below.\\  

However the most ambitious goal is to try to define a sort
of Floer homology using the Gabai moduli space. In order to do that
one needs to develop a Morse theory for foliated manifolds. One has at
 first to find a Morse function whose critical points will
be the taut codim-1 foliations. Immitating perhaps naively the Floer
homology case we have two natural candidates for a Morse function:
tangential Chern-Simons forms and Chern-Simons forms for cyclic
cohomology as developped by Quillen not very long ago in \cite{quillen} (that's
a noncommutative geometry tool). The hope is that by using the Gabai
moduli space one might have a chance to avoid the problems with
reducible connections (ie the ``bubbling phenomenon'', see
\cite{donald}) when trying to define Floer homology groups for
3-manifolds which are not homology 3-spheres. There are some more versions of 
Floer homology available but they need some extra structure: a $spin^{c}$ 
structure for the Seiberg-Witten version (and use of the monopole equation 
instead of the instanton equation), or a symplectic structure (as in the 
original Floer attempt) or a complex 
structure (as in the Oszvath-Szabo approach where one uses complex holomorphic 
curves instead of instantons).\\

As we shall see, some kind of Morse theory is also needed in order to define Hodge theory for 
tangential cohomology. In ordinary Morse 
theory, given a compact smooth manifold, one considers a real valued function 
(called the Morse function) defined on the manifold and under favourable cases 
one can reconstruct the homology of the manifold by using the flows of the 
critical points of the Morse function. In a would-be Morse theory for foliated 
manifolds one would like to reconstruct the homology of the space of leaves 
using a suitable Morse function, but it is currently unclear which homology of 
the 3 above is more suitable. Moreover the critical points should correspond to
 taut foliations in order to use the Gabai moduli space. We think the above 
challenge is fascinating. Some progress towards a Morse theory for foliated 
manifolds has been accomplished and it will be presented again in section 4 below.\\

Let us sum up the situation: The basic idea is to try to construct topological 
invariants like the Ray-Singer torsion, the Casson 
invariant, Floer Homology groups etc (along with their quantum field theoretic 
analogues-correlation functions) by replacing flat bundles with taut codim-1 
foliations. This will enable one to use noncommuattive geometric tools.
In order to make some progress towards any of the aforementioned tasks, there is something which semms essential:
noncommutative versions of Hodge theory.\\

\section{Hodge theory for noncommutative differential forms}

In this section, an analogue of the Hodge theorem will be proved for NC differential forms and as an immediate  corollary a 
\emph{NC free bosonic propagator} will be constructed.\\

Let us briefly recall that the "clasical" Hodge Theorem states that on every smooth, compact, Riemannian 
manifold (also assumed oriented), each de Rham cohomology class has a unique harmonic representative (namely 
the Laplace operator vanishes). As a consequence, every closed form can be written as the sum of an exact 
form plus a harmonic form and moreover every form can be written as the sum of a harmonic form plus an exact 
form plus a coexact form.\\

We follow Quillen (\cite{quillen}): Let $A$ be a complex, unital associative algebra (in our case at hand, this role will be played 
by the foliation $C^{*}$-algebra but the theorem can be proved in this slightly more general setting) and let
$$\Omega ^{n}A:=A\otimes _{\mathbb{C}}\bar{A}^{\otimes n},$$
for $n>0$, where
$$\bar{A}=A/\mathbb{C}$$
whereas
$$\Omega ^{n}A=0, n<0$$
and
$$\Omega ^{0}A=A.$$
Hence we get an identification
$$a_0da_1...da_n\leftrightarrow (a_0,a_1,...,a_n).$$
Then we also define
$$\Omega A=\oplus _n\Omega ^nA$$
which is the graded algebra (GA) of noncommutative differential forms over $A$, the multiplication being 
defined 
via
$$(a_0,a_1,...,a_n)(a_{n+1},a_{n+2},...,a_{k})=\sum_{i=0}^{k}(-1)^{k-i}(a_0,...,a_{i}a_{i+1},...,a_k)$$
for $k>n$. Moreover we define the differential $d:\Omega ^nA\rightarrow\Omega ^{n+1}A$ as follows:
$$d(a_0da_1...da_n)=da_0da_1...da_n$$
or in an equivalent notation
$$d(a_0,a_1,...,a_n)=(1,a_0,a_1,...,a_n)$$
and hence
$$d\Omega ^nA\simeq\bar{A}^{\otimes n+1}$$
for $n\geq 0$. Thus $(\Omega A,d)$ becomes a DGA.\\

On $\Omega A$, we can also define the Hochschild differential
$b:\Omega ^{n}A\rightarrow\Omega ^{n-1}A$
given by 
$$b(a_0,a_1,...,a_n)=\sum _{j=0}^{n-1}(-1)^{j}(a_0,a_1,...,a_{j}a_{j+1},...,a_n)+(-1)^{n}
(a_na_1,a_2,...,a_{n-1}).$$
Thus one has that
$$b(\omega da)=(-1)^{|\omega |}(\omega a-a\omega )=(-1)^{|\omega |}[\omega ,a]$$
and
$$b(a)=0,$$ 
where $|\omega |$ denotes the degree of the differential form $\omega$.\\

One also has the Karoubi operator (see \cite{kar}) which is a degree zero operator on $\Omega A$ given by
$$k:\Omega ^{n}A\rightarrow\Omega ^{n}A$$
where
$$k(\omega da)=(-1)^{|\omega |}(da)\omega $$
(for negative degrees it is given by the identity).\\

{\bf Lemma 1.} \textsl{One has the following relation:
$$bd+db=1-k.$$}
{\bf Proof:} One has
$$(bd+db)(\omega da)=b(d\omega da)+(-1)^{|\omega |}d[\omega ,a]=(-1)^{|\omega |+1}[d\omega ,a]+(-1)^{|\omega |}
d[\omega ,a]=$$
$$=[\omega ,da]=\omega da - (-1)^{|\omega |}(da)\omega.$$
$\square$\\

An immediate corollary of the above is that $k$ commutes both with $d$ and $b$, namely
$$bk=kb$$
and
$$dk=kd.$$
The above also shows that $k$ is homotopic to the identity with respect to either of the differentials 
$b$ or $d$.\\

One can formally see $d$ and $b$ as adjoint to each other, playing the roles of $d$ and $d^*$ respectively 
on the 
de Rham complex of a Riemannian manifold and call
$$bd+db=1-k$$
the \emph{NC Laplacian}. The natural thing to do next is to examine the spectrum of the NC Laplacian and focus 
on the zero eigenvalue. We have the following result:\\

{\bf Proposition 1.} \textsl{On $\Omega A$ one has the \emph{harmonic decomposition}
$$\Omega A=Ker(1-k)^2\oplus Im(1-k)^2,$$
where the generalised nullspace $Ker(1-k)^2$ is analogous to the space of harmonic forms}.\\

We can define the \emph{harmonic projection} $P$ to be the projection operator which is one on the first term 
of the harmonic decomposition and zero on the second; it is the spectral projection for $k$ associated to the 
eigenvalue 1. Hence the harmonic decomposition can be written
$$\Omega A=P\Omega A\oplus P^{\perp}\Omega A$$
where by definition
$$P^{\perp}=1-P$$
is the spectral projection of $k$ associated to the set of eigenvalues which are different from 1.\\

{\bf Proof:} The proof is based on the following technical Lemma:\\

{\bf Lemma 2.} \textsl{The Karoubi operator $k$ on $\Omega ^{n}A$ satisfies the polynomial relation
$$(k^n-1)(k^{n+1}-1)=0.$$}
{\bf Proof of Lemma 2:} We have
$$k(a_0da_1...da_n)=(-1)^{n-1}da_na_0da_1...da_{n-1}=$$
$$=(-1)^n a_nda_0...da_{n-1}+(-1)^{n-1}d(a_na_0)da_1...da_{n-1}$$
which in the second equivalent notation reads
$$k(a_0,a_1,...,a_n)=(-1)^n(a_n,a_0,...,a_{n-1})+(-1)^{n-1}(1,a_na_0,...,a_{n-1}).$$
Moreover
$$k(da_0da_1...da_n)=(-1)^nda_nda_0...da_{n-1}.$$  
In particular on $\Omega ^nA$ we have that $k^{n+1}d=d$.\\

Next we consider
$$k^j(a_0da_1...da_n)=(-1)^{j(n-1)}da_{n-j+1}...da_na_0da_1...da_{n-j}$$
for $0\leq j\leq n$. Hence
$$k^n(a_0da_1...da_n)=da_1...da_na_0=a_0da_1...da_n+[da_1...da_n,a_0]=$$
$$=a_0da_1...da_n+(-1)^nb(da_1...da_nda_0)$$
which yields
$$k^n=1+bk^nd$$
on $\Omega ^n A$. Then
$$k^{n+1}=k+bk^{n+1}d=k+bd$$
and using the definition of the NC Laplacian we get
$$k^{n+1}=1-db.$$
Thus from 
$$k^n=1+bk^nd$$
and
$$k^{n+1}=1-db$$
we obtain that $k$ on $\Omega ^n A$ satisfies the polynomial relation
$$(k^n-1)(k^{n+1}-1)=0.$$
$\square$\\
 
This polynomial relation implies that $k$ is invertible since the polynomial has constant term 1.\\

We returm to the proof of Proposition 1: Since an operator satisfies a polynomial equation, it gives rise 
to a direct sum decomposition into generalised eigenspaces corresponding to the distinct roots of the polynomial.\\

The roots of 
$$(k^n-1)(k^{n+1}-1)$$
are the $n$ different n-th roots of unity and the $n+1$ different roots of unity of order dividing $n+1$. 
Yet $n$ and $n+1$ are relatively prime which means that these two sets of roots have only $k=1$ in common. 
Hence 1 is a double root and and all other roots are simple.\\

Consequently $\Omega ^nA$ decomposes into the direct sum of the generalised eigenspace $Ker(1-k)^2$ 
corresponding to the eigenvalue $z=1$ and the ordinary eigenspaces $Ker(k-z)$ for each root of unity $z\neq 1$ 
of order dividing $n$ or $n+1$.\\

Combining the above $\forall n$ we obtain the following spectral decomposition with respect to $k$
$$\Omega A =Ker (1-k)^2 \oplus [\bigoplus _{z\neq 1}Ker (k-z )].$$
Lumping the eigenvalues $z\neq 1$ together we have
$$\Omega A=Ker(1-k)^2\oplus Im(1-k)^2$$
which completes the proof.\\

$\square$\\

Note however that the NC Laplacian
$$1-k=[b,d],$$
contrary to the Riemannian manifold situation, \textsl{is only nilpotent on the first factor (and invertible 
on the second)}. This defect can be cured by introducing the \emph{rescaled NC Laplacian}
$$L=[b,Nd]$$
where $N$ is the numbering operator (this is a degree zero operator which acting on forms gives the scalar 
multiple of the form by its degree). \emph{The rescaled NC Laplacian then vanishes on $P\Omega A$ (and is 
invertible on its complement)}.\\

On the complementary space $P^{\perp}\Omega A$ the NC Laplacain is invertible and homotopic to zero with 
respect to either differential $b$ or $d$. Thus we can define the \emph{Green's operator} $G$ for the NC 
Laplacian which is equal to its inverse on $P^{\perp}\Omega A$, namely
$$G=(1-k)^{-1}$$
and $G=0$ on $P\Omega A$. This can be seen as the \emph{NC free bosonic propagator}.\\

As in the clasical Hodge theory, the complementary space $P^{\perp}\Omega A$ to the "harmonic fomrs" splits 
into subspaces of exact and coexact forms:\\

{\bf Proposition 2.} \textsl{One has}
$$P^{\perp}\Omega A=dP\Omega A\oplus bP\Omega A.$$

{\bf Proof:} This is a formal cosnequence of the identity
$$G(bd+db)=1$$
on $P^{\perp}\Omega A$ and the fact that $G$ commutes with both differentials $b$,$d$ (as one can check via a 
direct computation). Thus
$$(Gdb)d=G(bd+db)d=d$$
implies that $Gdb$ is a projection with image $dP^{\perp}\Omega A$. Similarly $Gbd$ is a projection with 
image $bP^{\perp}\Omega A$ and as these projections add to 1 we get the desired decomposition.\\

$\square$\\ 

Let us close this section with some comments: There are 5 basic TQFT's known up to now: The (2+1) 
Abelian Chern-Simons theory due to Albert Schwarcz, its non-Abelian generalisation (this is the so-called 
Jones-Witten theory), the (3+1) Donaldson-Floer -Witten theory, its dual (the so-called Seiberg-Witten theory) 
and the Kontsevich-Gromov-Witten theory (topological $\sigma$ models) and their generalisations.\\

The simplest of all is the Abelian Chern Simons theory where the Lagrangian density is given by the Abelian 
Chern-Simons 3-form
$$S=\int_{N^3}A\wedge dA$$
and the partition function is given by the following product of zeta-function regularised determinants of 
Laplacians
$$Z(N^3)=(Det_{\zeta}\Delta _{1})^{-1/4}(Det _{\zeta}\Delta _{0})^{3/4}.$$

\emph{The NC version of this should be obtained in a straightforward way for the case of a, say, 
noncommutative 3-sphere and using the NC Laplacian. We hope to be able to report on this elsewhere} 
(see \cite{quillen} for Chern-Simons forms in NCG and \cite{connes3} and \cite{connes4} for 
noncommutative 3-spheres).\\

\section{Hodge Theory for Tangential Cohomology}

Let $(M,F)$ be a smooth foliation on a closed $n$-manifold $M$ (and $F$ is an integrable subbundle of the 
tangent bundle $TM$ of $M$ where $dimF=p$, $codimF=q$ with $p+q=n$), equipped with a holonomy invariant transverse measure
$\Lambda$ (we need that in order to be able to perform the analogue of "integration along the fibres" which we do for vector or 
principal
$G$-bundles using the Haar measure which is invariant under the group action). We consider the \emph{tangential cohomology} coming 
from the 
differential graded complex $(d_{F},\Omega ^{*}(M,F))$, where $d_F$ denotes the \emph{tangential} exterior derivative (namely taking 
derivatives only
along the tangential (leaf) directions) and $\Omega ^{*}(M,F)$ denotes forms on $M$ with values on the bundle $F$. Due to the 
integrability of $F$, the tangential exterior derivative is also a differential, namely
$d_{F}^{2}=0$, hence we can take the cohomology of the above complex. We pick a Riemannian metric $g$ on $M$ (which, when restricted 
to every leaf gives
a Riemmanian metric on every leaf), we consider the adjoint operator $d_{F}^{*}$ and we form the \emph{tangential Laplacian} 
$\Delta _{F}:=d_{F}^{*}d_{F}+d_{F}d_{F}^{*}$.
We denote by $\beta _{k}$ the $k$-th tangential Betti number ($0\leq k\leq p$), where clearly 
$\beta _{k}=dim_{\Lambda}[Ker (\Delta _{F}^{k})]$. 

[Aside 5: We must make an important remark here: this is the Murray-von Neumann dimension 
defined by Connes using 
the invariant transverse measure, it is finite;
the tangential cohomology groups may be infinite dimensional as linear spaces (see \cite{moore})].\\

It is well known that  there exist real valued smooth functions on $M$ having only Morse or birth - death singularities. 
We shall denote by $h$ and $v$) cummulatively the horizontal (or tangential) and vertical (or transverse) local coordinates 
respectively and by 
$L_x$ the leaf through the point $x \in M$. 
For any smooth real function $\phi$  on $M$ we denote by $d_{F}\phi $  the differential of $\phi$ in the leaf (horizontal or tangential) 
directions. A 
point $a \in M$ for which the leaf differential vanishes will be called a \emph{tangential singularity} for $\phi$. For such a 
singularity the 
horizontal (or tangential) Hessian $d^{2}_{F} \phi $ makes sense and in local coordinates $(h,v)$ one has

$$d_{F} \phi  (h,v) = \sum _{1 \leq i \leq p} \frac{\partial\phi }{\partial h^{i}} (h,v)$$

and 

$$d^{2}_{F}\phi (h,v)=(\frac{\partial ^{2}\phi }{\partial h^{i} \partial h^{j}}(h,v))_{ij}$$

The index of a tangential singularity $a$ on $M$ is defined as the number of minus signs in the signature of the quadratic form  
$d^{2}_{F}\phi (a)$.\\

{\bf Definition 1:} A tangential singularity $a$ on $M$ of a smooth real function $\phi$  on $M$ as above is called a  
\textsl{Morse singularity} if $d^{2}_{F} \phi (a)$ is non-singular.\\

We denote by $T(\phi )$ (resp $M(\phi ), M_{i}(\phi ))$ the set of all tangential singularities (resp. of Morse singularities, 
Morse singularities of index $i$, where $0\leq i\leq p$) of the function $\phi $.  The first complication emerges since in this case 
a good definition for a tangential (or horizontal) Morse function cannot be reduced to simply a smooth function on $M$ having only 
tangential Morse singularities. This is explained in the following Lemma:\\

{\bf Lemma 3:} Let $(M,F)$ be as above. Assume that there exists a smooth function $\phi$  again as above with only tangential Morse 
singularities. Then the set of all tangential Morse singularities of $\phi$  is a closed $q$-dim submanifold transverse to the foliation.\\

{\bf Proof:} We suppose that $\phi$ is a smooth function on $M$ such that for any leaf $L$ in the quotient space $M/F$ the restriction 
$\phi |_{L}$
of $\phi$  on the leaf $L$ has no degenerate critical points. Then the map

$$x\mapsto (x,d_{F}\phi (x))$$

from $M$ to $T^{*}F$ is transverse to the zero section of $T^{*}F$ since its differential is given on any foliation chart 
$\Omega = U \times T$ by

$$d(d_{F}\phi )(h,v)(X_h,X_v) = ((X_h,X_v),\phi _{hh}(h,v)X_h + \phi _{hv}(h,v)X_v)$$

where the subscripts  $h$, $v$ denote partial derivative with respect to the corresponding coordinates and $det(\phi _{hh}(h,v))\neq 0$ 
for a Morse singularity with coordinates $(h,v)$. This implies first that the set of all Morse singularities   $M(\phi )$ is a closed 
submanifold of $M$ with $dim(M(\phi )) = codim(F)$ and second that  $M(\phi )$ is transverse to the foliation $F$ because for any 
non-zero tangent vector $X = (X_h,X_v)$ of $M(\phi )$ at the point $(h,v)$, one has that

$$\phi _{hh}(h,v)X_h + \phi _{hv}(h,v)X_v = 0.$$

This means that the transverse component $X_v$ of $X \neq 0$ is non-zero which proves that $M(\phi )$ is transverse to the foliation 
$F$. This concludes the proof.

$\square$\\

It turns out that many interesting foliations have no closed transversals and hence any good notion of tangential Morse function should 
allow degenerate critical points in the leaf direction. (However taut foliations which are the ones appearing in the Gabai moduli 
space do have closed transversals by definition).\\

{\bf Definition 2:} We call \textsl{almost Morse function} a smooth function $\phi$ as above with degenerate 
critical points which only occure at a $\Lambda$-negligible set of leaves (namely we allow degenerate critical points but not too many).\\

{\bf Definition 3:} A \textsl{good} almost Morse function is an almost Morse function which is generically unfolded in the sense of Igusa 
Parametrised Morse Theory (see \cite{igusa}) 
(roughly this means that it has only birth-death singularities, namely points where critical points cancel or create in pairs). More 
concretely, the last 
requirement means that there exist 
\textsl{normal forms}
describing the function in a neighbourhood of a birth-death singularity. A birth-death singularity is a degenerate 
tangential singularity (i.e. tangential Hessian 
vanishes) for which the restriction of the map $x\mapsto (d_{F}(\phi)(x),det[d_{F}^{2}(\phi)(x)])$ has rank $p$ at $x$.\\

\subsection{Witten's perturbation by a Morse function-Tangential version}

Let $(M,F)$ be a foliation as above equipped with a holonomy invariant transverse measure $\Lambda$. We choose a smooth Riemannian metric 
on $M$ and denote by $\Delta ^{k}_{L}$ $(0 \leq k \leq p)$ the corresponding Laplace operator on the leaf $L$ acting on $k$-forms. 
We know that the bundle of 
Hilbert spaces is square integrable and thus has a well-defined Murray-von Neumann dimension

$$\beta _{k} = dim_{\Lambda} (Ker (\Delta ^{k}_{L})) < \infty$$

which does not depend on the choice of metric. Assume moreover that $codim(F) \leq dim(F)$.
Let $\phi$ be a smooth real function on $M$ which is good almost Morse function and $\tau$ a positive real parameter. For each leaf 
$L$ and $0 \leq k \leq p$ we denote by $d^{k}_{\tau ,L}$ the closure (in $L^{2} (L,\wedge ^{*}F)$, the space of square integrable forms 
on the 
leaf $L$) of the 
operator which sends each smooth $k$-form $\omega$ on $L$ to the smooth $(k+1)$-form $e^{-\tau\phi} d^{k}_{L}(e^{\tau\phi}\omega)$ 
again on $L$.\\

{\bf Definition 4:} We shall call \textsl{Witten tangential Laplacian} the measurable filed $(\Delta ^{k}_{\tau ,L})_{L}$ which is 
defined in the obvious way, namely

$$\Delta ^{k}_{\tau ,L} = d^{k-1}_{\tau ,L}(d^{k-1}_{\tau ,L})^{*} + (d^{k}_{\tau ,L})^{*}d^{k}_{\tau ,L}.$$

 Then we 
prove that $\Delta ^{k}_{\tau}$ computes the ($L^{2}$) tangential cohomology of $(M,F)$:\\

{\bf Proposition 3:} The fields $(Ker (\Delta ^{k}_{\tau ,L} ))_{L}$ and  $(Ker (\Delta ^{k}_{L}))_{L}$ of Hilbert spaces are 
measurably isomorphic and one has that

$$\beta _{k} = dim_{\Lambda} [Ker (\Delta ^{k}_{\tau ,L})_{L}] < + \infty$$

for any positive real $\tau$ and $0 \leq k \leq p$.\\

{\bf Proof:} The proposition can be proved following the steps below:\\

1.The operator $d^{k}_{\tau} = (d^{k}_{\tau ,L})_{L}$ is a differential operator which is elliptic along 
the leaves of $F$. 
This can be proved using an argument similar to the one used by Connes in \cite{connes} to prove the transversal index theorem.\\

2. Observe that the adjoint  of $d^{k}_{\tau ,L}$ is the closure of the operator which sends each smooth $(k+1)$-form 
$\omega$ on $L$ 
to the smooth $k$-form $e^{\tau\phi} (d^{k}_{L})^{*}(e^{-\tau\phi})\omega$ again on $L$, where

$$(d^{k}_{L})^{*} = (-1)^{pk+1} *d^{k}_{L}* ,$$ 

and where "$*$" denotes the Hodge star operator on the leaf $L$ defined via the Remanian metric.\\

3. We note that $\Delta ^{k}_{\tau} = (\Delta ^{k}_{\tau ,L})_{L}$  is a field of measurable positive operators acting on the 
Hilbert space of 
square integrable $k$-forms on the leaf $L$. Moreover $\Delta ^{k}_{\tau}$  is elliptic along the leaves.\\

4. For any leaf $L$ and $0 \leq k \leq p$, we denote by $T^{k}_{L}$ the bounded operator on $L^{2}(L, \wedge ^{k}T^{*}F)$ defined by

$$T^{k}_{L}(\omega )(x) = e^{-\tau\phi (x)}\omega (x),$$  

for $\omega \in L^{2}(L, \wedge ^{k}T^{*}F)$.\\

It is clear that $T^{k}_{L}$ is invertible and defines an element of  $L^{\infty}(M/F, \wedge ^{k}T^{*}F)$. Next we set 

$$U^{k}_{\tau ,L} = Q^{k}_{\tau ,L} T^{k}_{L} Q^{k}_{L},$$

where $Q^{k}_{\tau ,L}$ (resp. $Q^{k}_{L}$) denotes the orthogonal projection onto the subspace $Ker (\Delta ^{k}_{\tau ,L})$ 
(resp. onto $Ker (\Delta ^{k}_{L}))$. We thus define 
a measurable field $(U^{k}_{\tau ,L})_{L}$ of endomorphisms of the random Hilbert space $(L^{2}(L, \wedge ^{k}T^{*}F))_{L}$, such that 
$Ker (\Delta ^{k}_{\tau ,L})$ is a superset of  $U^{k}_{\tau ,L}(Ker (\Delta ^{k}_{L}))$.

We want to show that $(U^{k}_{\tau ,L})_{L}$ belongs to $L^{\infty}(M/F, \wedge ^{k}T^{*}F)$ and defines an isomorphism of Hilbert spaces 
from $(Ker (\Delta ^{k}_{L}))_{L}$ to $(Ker (\Delta ^{k}_{\tau ,L}))_{L}$. One then has (omitting the subscript L):

$$d^{k}_{\tau}  = T^{k+1}_{\tau} d^{k}(T^{k}_{\tau} )^{-1}$$

and hence

$$T^{k}_{\tau} (Ker d^{k}) = Ker d^{k}_{\tau} (equation 1)$$

$$T^{k+1}_{\tau}[cl.Im(d^{k})] = cl.Im(d^{k}_{\tau}) (equation 2).$$

But it follows from Hodge theory that one has the following orthogonal decompositions:

$$Ker(d^k) = Ker(\Delta ^{k}) \oplus cl.Im(d^{k-1})$$

and

$$Ker(d^{k}_{\tau}) = Ker(\Delta ^{k}_{\tau}) \oplus cl.Im(d^{k-1}_{\tau}),$$

and then from equations (1) and (2) it follows that $T^{k}_{\tau}$ is given in those decompositions by a $2\times 2$ matrix with the upper 
left entry being $U^{k}_{\tau}$, the lower right entry being $B^{k}_{\tau}$, the upper right entry being $0$ and the lower left entry 
being any 
element, namely 

\[
T^{k}_{\tau}=
\left( {\begin{array}{cc}
U^{k}_{\tau} & 0 \\
* & B^{k}_{\tau} \\
\end{array} } \right)
\]

and where the entry $B^{k}_{\tau ,L} = T^{k}_{\tau ,L} | cl.Im(d^{k-1}_{L})$, (namely $T^{k}_{\tau ,L}$ restricted to 
$cl.Im(d^{k-1}_{L})$, 
the closure of the 
Image of $d^{k-1}_{L}$), is invertible. We thus deduce that $U^{k}_{\tau}$ is an isomorphism from $(Ker (\Delta ^{k}_{L}))_{L}$ onto 
$(Ker (\Delta ^{k}_{\tau ,L}))_{L}$ and hence

$$\beta _{k} = dim_{\Lambda} [Ker (\Delta ^{k}_{L})_{L}] = dim_{\Lambda} [Ker (\Delta ^{k}_{\tau ,L})_{L}]$$

and this holds $\forall\tau > 0$. As $\beta _{k} < + \infty$, then an argument similar to  Connes \cite{connes} completes  the proof.
$\square$

[Aside 6: Connes and Fack have proved 
that \textsl{every} measured foliation with 
$q\leq p$ has 
at least one good tangential almost Morse function; their proof is based on an astounding theorem due to K. Igusa: it 
was a well-known fact that a generic smooth
real valued function on a closed manifold has only nondegenerate critical points; however a generic 1-parameter family
 of real valued smooth functions has in addition
birth-death points where critical points are created or canceled in pairs. A multi-parameter family has a zoo of 
complicated singularities; K. Igusa proved that more 
complicated singularities can be avoided: for any foliation on a closed manifold it is always possible to find a
 smooth real valued function such that singularities
 associated with the critical points of its restriction to every leaf are at most of degree 3! Clearly we think of a 
foliation as a more complicated parametrised family
of manifolds than a fibre bundle: the family of manifolds (leaves-they correspond to the tangential directions) is 
parametrised by the space of leaves 
(corresponds to the transverse directions); in a fibre bundle we have a family of manifolds (fibre) parametrised by the
 base manifold].\\

{\bf Comments:}

It is not true that any measured foliation with $q\leq p$ has a tangential Morse function, namely the foliations with
 tangential Morse functions are rather
special (they must have a closed transversal); taut foliations nevertheless, which is what we are mostly interested in,
 do have, by definition, a complete
closed transversal).\\

If we denote by $A(M,F)$, $J(M,F)$ and $R(M,F)$ the sets of tangential almost Morse functions, tangential generalised 
Morse functions and 
tangential generalised Morse functions which are generically unfolded respectively, then the good tangential almost
 Morse functions are those in the intersection of
$A(M,F)$ and $R(M,F)$ (clearly the 3rd set is a subset of the second). The hard piece due to K. Igusa is to prove that
 for a closed $M$ and an $F$ with
$codimF\leq dimF$, the set $J(M,F)$ is nonempty.

For $\phi$ a good tangential almost Morse function, we have that the critical manifold $S_{F}(\phi)$ is a $q$-dim 
submanifold of $M$ transverse to $F$, the set of 
tangential Morse singularities of index $i$ $S_{1,F}^{i}(\phi)$ is also a $q$-submanifold of $M$ transverse to $F$ and 
open inside the critical manifold (but not closed
in $M$ in general) and the set of tangential birth-death singularities $S_{2,F}^{i}(\phi)$ of index $i$ of $\phi$ is a 
closed $(q-1)$-submanifold of the critical manifold
and it is both in the closure of $S_{1,F}^{i}(\phi)$ and of $S_{1,F}^{i+1}(\phi)$.\\

\emph{Acknowledgements:} 
The author wishes to thank IHES for its hospitality and for providing a very stimulating scientific environment, 
a magnificent natural environment along with excellent working facilities. Moreover we would like to thank 
Alain Connes, Tierry Fack, Kim Froysov, David Gabai for useful discussions and communication and J.L. Heitsch for providing a copy of 
his joint article with C. Lazarov.  Furthermore the author  has also been benefited from graduate 
lecture courses on Floer Homology by Simon Donaldson, Peter Kronheimer and others during his Oxford days.\\


\begin{thebibliography}{20}




\bibitem{atiyah}M. F. Atiyah: \emph{"Topological Quantum Field Theories"}, IHES Publ. Math. 68 (1989), 175-186.\\

M. F. Atiyah: \emph{"The Geometry and Physics of Knots"}, Accademia Nazionale dei Lincei, Cambridge University Press 1990.\\

\bibitem{connes} A. Connes: \emph{``Noncommutative Geometry''}, Academic Press 1994.\\



\bibitem{geo600} C.J. Hogan: \emph{Measurement of Quantum Fluctuations in Geometry}, Phys. Rev. D, Vol 77 (2008).\\


\bibitem{connes2} \textsl{A.H. Chamseddine, A. Connes, M. Marcolli}: \emph{"Gravity and the Standard Model 
with Neutrino Mixing"}, Adv. Theor. Math. Phys. 11 (2007), 991-1089.\\ 

\bibitem{connes3} \textsl{A. Connes, M. Dubois-Violette}: \emph{"Moduli Space and Structure of Noncommutative 
3-spheres"}, Lett. Math. Phys. 66 (2003), 91-121.\\ 

\bibitem{connes4} \textsl{A. Connes, M. Dubois-Violette}: \emph{"Non commutative finite dimensional manifolds 
II. Moduli space and structure of non commutative 3-spheres"}, math/0511337.\\ 


\bibitem{floer}A. Floer: \emph{"An instanton invariant for 3-manifolds"}, Commun. Math. Phys. (1989).\\

\bibitem{thooft}G. 't Hooft: \emph{``Dimensional Reduction in Quantum Gravity''},
Essay dedicated to Abdus Salam in \textsl{``Salamfestschift: a collection of
talks''}, editors: A. Ali, J. Ellis and S. Randjbar-Daemi, World Scientific
1993 and gr-qc/9310026.\\

\bibitem{cds} A. Connes, M.R. Douglas and A. Schwarz: 
\emph{``Noncommutative geometry and Matrix theory: compactification on tori''} 
JHEP02(1998)003.\\

\bibitem{sw} N.Seiberg, E.Witten: \emph{"String theory and Noncommutative Geometry"}, JHEP09(1999)032.\\

\bibitem{witten} E. Witten: \emph{"Quantum Field Theory and the Jones Polynomial"}, Commun. Math. Phys. 121 
(1989), 353-386.\\

E. Witten: \emph{"Supersymmetry and Morse Theory"}, J. Diff. Geom. 17 (1982).\\

\bibitem{donald} S.K. Donaldson: \emph{``Floer Homology Groups in Yang-Mills
Theory''}, Cambridge University Press 2002.\\

S.K. Donaldson and P.B. Kronheimer: \emph{``The geometry of 4-manifolds''}, 
Oxford University Press 1991.\\

\bibitem{rs}D. B. Ray and I. M. Singer: \emph{"R-torsion and the Laplacian on Riemannian manifolds"}, 
Adv. Math. 7 (1971) 145-210.\\

\bibitem{kar} \textsl{M. Karoubi}: \emph{"Homologie cyclique et K-Theorie"}, Asterisque 149 (1987).\\


\bibitem{luck} W. Luck: \emph{``$L^{2}$-invariants: Theory and Applications to 
Geometry and K-Theory''}, A Series of Modern Surveys in Mathematics Vol 44, 
Springer, 2002.\\

\bibitem{moore} C.C. Moore and C. Schochet: \emph{``Global Analysis on Foliated 
Manifolds''}, Springer (1988).\\

\bibitem{gabai} D. Gabai: \emph{''Essential Laminations and Kneser Normal Form''},
Jour. Diff. Geom. Vol 53 No 3 (1999).\\

\bibitem{zois} I.P. Zois: \emph{``A new invariant for $\sigma $-models''}, Commun.
Math. Phys. Vol 209 No 3 (2000).\\

\bibitem{candel} A. Candel and L. Conlon: \emph{``Foliations I and II''},
Graduate Studies in Mathematics Vol 23, AMS, Oxford University Press
(2000) and (2003).  \\ 

\bibitem{quillen} D.G. Quillen: \emph{"Chern-Simons Forms and Cyclic Cohomology"}, in \emph{``The Interface of 
Mathematics and Particle Physics''}, editors D.G. Quillen, G.B. Segal and S.T. Tsou, Oxford University Press 1990.\\ 

J. Cuntz and D.G. Quillen: \emph{``Operators on noncommutative differential forms and
 cyclic homology''} in \textsl{``Geometry, Topology and Physics for R. Bott''}, 
International Press 1995, edited by S-T Yau.\\

\bibitem{heitsch} J.L. Heitsch and C. Lazarov: \emph{``Riemann-Roch-Grothendieck and 
Torsion for Foliations''} (unpublished).\\

\bibitem{bismut} J.-M. Bismut and J. Lott: \emph{``Flat Vector Bundles, Direct Images and Higher Real Analytic Torsion''}, 
J. Am. Math. Soc. 8 (1995).\\

\bibitem{igusa}K. Igusa: \emph{"Higher singularities of smooth functions are unnecessary",} Annals of Math. 119, 
(1984), 1-58.\\

K. Igusa: \emph{"Parametrised Morse Theory and its Applications"}, Proc. Internat. Congr. Math. (Kyoto 1990), 
Math. Soc. Japan, Tokyo, (1991), 643-651.\\
\end{thebibliography}
\end{document}